\documentclass[aps,prd,notitlepage,showpacs,nofootinbib,superscriptaddress]{revtex4-1}

\usepackage{graphicx}
\usepackage[utf8]{inputenc} 

\usepackage{amsmath}
\usepackage{amssymb}
\usepackage{float}
\usepackage{comment}
\usepackage{slashed}
\usepackage[normalem]{ulem}

\usepackage{caption}
\usepackage{subcaption}
\usepackage{amsmath}
\usepackage{amssymb}

\usepackage[usenames,dvipsnames]{color}
\usepackage[colorinlistoftodos]{todonotes}
\usepackage[colorlinks=true,citecolor=darkred,urlcolor=darkred, pdfborder={0 0 0}]{hyperref}
\usepackage[normalem]{ulem}

\definecolor{darkred}{rgb}{0.6,0,0}
\usepackage[colorinlistoftodos]{todonotes}

\definecolor{linkcolor}{rgb}{0,0,0.5}
 
\newcommand {\ignore}[1]{}

\def \znbb {$\rm 0\nu\beta\beta$ }

\def\gsim{\raise0.3ex\hbox{$\;>$\kern-0.75em\raise-1.1ex\hbox{$\sim\;$}}}
\def\lsim{\raise0.3ex\hbox{$\;<$\kern-0.75em\raise-1.1ex\hbox{$\sim\;$}}}

\def\SM{$\mathrm{SU(3)_c \otimes SU(2)_L \otimes U(1)_Y}$ }
\newcommand{\sm}{{Standard Model }}

\providecommand{\be}{ \begin{equation} } 
\providecommand{\ee}{ \end{equation} }
\providecommand{\bea}{\begin{eqnarray}}
\providecommand{\eea}{\end{eqnarray}}
\providecommand{\nn}{\nonumber}

\usepackage{soul}

\definecolor{mightnightblue}{RGB}{25,25,112}

\definecolor{brown}{rgb}{0.59, 0.29, 0.0}

\def\vev#1{\left\langle #1\right\rangle}
\def\SM{$\mathrm{SU(3)_c \otimes SU(2)_L \otimes U(1)_Y}$ }
\def\21{$\mathrm{SU(2)_L \otimes U(1)_Y}$}

\def\sm{standard model }

\def\3311{$\mathrm{SU(3) \otimes SU(3)_L \otimes U(1)_X \otimes U(1)_{N}}$ }

\newcommand{\AddrAHEP}{%
  AHEP Group, Institut de F\'{i}sica Corpuscular --
  C.S.I.C./Universitat de Val\`{e}ncia, Parc Cient\'ific de Paterna.\\
 C/ Catedr\'atico Jos\'e Beltr\'an, 2 E-46980 Paterna (Valencia) - SPAIN}

\begin{document}

\title{\boldmath \color{BrickRed} A theory for scotogenic dark matter stabilised by residual gauge symmetry}

\author{ Julio Leite }\email{julio.leite@ific.uv.es}
\affiliation{\AddrAHEP}
\affiliation{Centro de Ci\^encias Naturais e Humanas, Universidade Federal do ABC,\\
09210-580, Santo Andr\'e-SP, Brasil}

\author{Oleg Popov} \email{opopo001@ucr.edu}
\affiliation{Institute of Convergence Fundamental Studies, Seoul National University of Science and Technology, \\Seoul 139-743, Republic of Korea }
\affiliation{Department of Physics, Korea Advanced Institute of Science and Technology, \\291 Daehak-ro, Yuseong-gu, Daejeon 34141, Republic of Korea}
\author{Rahul Srivastava}\email{rahul@iiserb.ac.in}
\affiliation{\AddrAHEP}
\affiliation{India Institute of Science Education and Research - Bhopal, Bhopal Bypass Road, Bhauri, 462066, Bhopal, India}
\author{Jos\'{e} W. F. Valle}\email{valle@ific.uv.es}
\affiliation{\AddrAHEP}

\begin{abstract}
\vspace{0.5cm}

Dark matter stability can result from a residual matter-parity symmetry, following naturally from the spontaneous breaking of the gauge symmetry.
Here we explore this idea in the context of the $\mathrm{SU(3)_c \otimes SU(3)_L \otimes U(1)_X \otimes U(1)_{N}}$ electroweak extension of the standard model.
The key feature of our new scotogenic dark matter theory is the use of a triplet scalar boson with anti-symmetric Yukawa couplings.
This naturally implies that one of the light neutrinos is massless and, as a result, there is a lower bound for the \znbb decay rate.

\end{abstract}

\maketitle
\noindent

\section{Introduction}
\label{Sect:intro}

In order to account for the existence of cosmological dark matter, we need new particles not present in the Standard Model (SM) of particle physics.
Moreover, new symmetries capable of stabilising the corresponding candidate particle on cosmological scales are also required.
Here we focus on the so-called Weakly Interacting Massive Particles, or WIMPs, as dark matter candidates. 
Within supersymmetric schemes, WIMP stability follows from having a conserved R-parity symmetry~\cite{Jungman:1995df}. 
Our present construction does not rely on supersymmetry nor on the imposition of any \textit{ad hoc} symmetry to stabilise dark matter.
It is also a more complete theory setup, in the sense that it naturally generates neutrino masses as well.
These arise radiatively, thanks to the exchange of new particles in the ``dark'' sector.
The procedure is very well-motivated since neutrino masses are anyways necessary to account for neutrino oscillation data~\cite{deSalas:2017kay}.

Here we follow an alternative approach that naturally incorporates neutrino mass right from the beginning. 
This is provided by scotogenic dark matter schemes.
These are ``low-scale'' models of neutrino mass~\cite{Boucenna:2014zba} where dark matter emerges as a radiative mediator of neutrino mass generation.
In this case, the symmetry stabilising dark matter is also responsible for the radiative origin of neutrino masses in a very elegant way~\cite{Ma:2006km}. 
Yet, in this case too, a dark matter stabilisation symmetry is introduced in an \textit{ad hoc} manner.
The need for such ``dark'' symmetry is a generic feature also of other scotogenic schemes, such as the generalization proposed in~\cite{Hirsch:2013ola,Merle:2016scw}. 

Extending the \SM gauge symmetry can provide a natural setting for a theory of dark matter where stabilisation can be automatic~\cite{Alves:2016fqe,Dong:2017zxo, Kang:2019sab}.
Such electroweak extensions involve the SU(3)$_{\rm L}$ gauge symmetry, which also provides an ``explanation'' of the number of quark and lepton families from the anomaly cancellation requirement~\cite{Singer:1980sw,Pisano:1991ee,Frampton:1992wt}. 
For recent papers using the  SU(3)$_{\rm L}$ gauge symmetry see 
Refs.~\cite{Hernandez:2013mcf,Boucenna:2014ela,Hernandez:2014lpa,Queiroz:2016gif,Addazi:2016xuh,CarcamoHernandez:2017cwi,Barreto:2017xix,Dong:2018aak,Dias:2018ddy,Huitu:2019bvo}. 
These theories can also, in some cases, be made consistent with unification of the gauge couplings~\cite{Boucenna:2014dia,Deppisch:2016jzl} 
and/or with the existence of left-right gauge symmetry~\cite{Reig:2016tuk,Hati:2017aez}. 
In the extended electroweak gauge symmetry models discussed in~\cite{Alves:2016fqe,Dong:2017zxo} the stability of dark matter results from the presence of a matter-parity symmetry, $M_P$, a non-supersymmetric version of R-parity, that is a natural consequence of the spontaneous breaking of the extended gauge symmetry.

The purpose of this letter is to improve upon the proposal in~\cite{Kang:2019sab} in two ways.
First, we simplify the particle content. Compared with Ref.~\cite{Kang:2019sab} no extra vector-like fermions nor scalar SU(3)$_{\rm L}$ sextets are needed. 
Instead, the matter parity odd, third component ($N_L$) of the SU(3)$_{\rm L}$ lepton triplet  plays the role of dark fermion with its Dirac partner being a new SU(3)$_{\rm L}$ singlet
$N_R$. Moreover, the dark sextet scalar particles are replaced by an SU(3)$_{\rm L}$ scalar triplet. As seen in Fig.~\ref{fig1} these particles are enough to implement the scotogenic scenario.  
Moreover, in contrast to the proposal in Ref.~\cite{Kang:2019sab}, here we predict that one of the light neutrinos is massless. This feature arises in a novel way when compared to other schemes in the literature. So far most realistic theories where one of the neutrinos is (nearly) massless typically involve ``missing partner'' schemes, such as the ``incomplete'' seesaw mechanism~\cite{Schechter:1980gr} or similar radiative mechanisms~\cite{Reig:2018ztc}. 

The paper is organized as follows. 
In Sec.~\ref{sec:model} we present the model, while the loop-induced neutrino masses are discussed in Sec~\ref{sec:neutrino-masses}. 
The symmetry breaking sector, scalar potential and mass spectrum are discussed in Sec.~\ref{sec:scalar-sector}. 
Concerning phenomenology, in Secs. \ref{sec:0nbb} and \ref{sec:pheno}, we briefly comment on dark matter and the predicted lower bound for
the \znbb decay rate, as well as FCNC and collider signatures. Finally, our conclusions are presented in Sec. \ref{sec:Conclusions}.


\section{Our model}
\label{sec:model}


Here we give the main features of the model, based on the $\mathrm{SU(3)_c \otimes SU(3)_L \otimes U(1)_X \otimes U(1)_{N}}$ gauge invariance. 
The electric charge and $B-L$ generators are given by 
\bea\label{QBL}
    Q & = & T_3 - \frac{1}{\sqrt{3}} T_8 + X~,\\
    B-L & = & -\frac{2}{\sqrt{3}} T_8 + N~,
\eea
where $T_m$, with $m = 1,2,3,...,8$, are the $\mathrm{SU(3)_L}$ generators, whereas $X$ and $N$ are the $\mathrm{U(1)_X}$ and $\mathrm{U(1)_N}$ generators, respectively. 
Notice that, due to the extra $U(1)_N$ symmetry, the $B-L$ symmetry is fully gauged. 
The SM $\mathrm{SU(2)_L}$ doublet quarks and leptons reside inside the $SU(3)_L$ anti-triplet $q_{iL}$, triplet $q_{3L}$; $i=1,2$ and $l_{aL}$; $a= 1,2,3$ and their field decomposition is given by:
\begin{eqnarray}
 q_{iL} & = & \begin{pmatrix} d_{iL} \\ -u_{iL} \\ D_{iL} \end{pmatrix} \, , \qquad 
 q_{3L} \, = \, \begin{pmatrix} u_{3L} \\ d_{3L} \\ U_{3L} \end{pmatrix} \, , \qquad 
 l_{aL} \, = \, \begin{pmatrix} \nu_{aL} \\ e_{aL} \\ N_{aL} \end{pmatrix} \, ,
 \label{eq:field-decomposition}
\end{eqnarray}
whereas their $\mathrm{SU(2)_L}$ singlet partners are given by $u_{aR}$, $d_{aR}$ and $e_{aR}$ respectively.
The full particle content of the model along with the corresponding charges is summarised in Table \ref{tab}. 

Symmetry breaking takes place through the non-vanishing vacuum expectation values (vevs) as given below,
\bea\label{vevs}
\vev{\sigma}&=&\frac{v_\sigma}{\sqrt{2}}~,\quad\langle\chi\rangle=\frac{1}{\sqrt{2}}(0,0,w)^T~,\\
\vev{\eta}&=&\frac{1}{\sqrt{2}}(v_1,0,0)^T,\quad \langle\rho\rangle=\frac{1}{\sqrt{2}}(0,v_2,0)^T,\quad
\vev{\zeta}=\frac{1}{\sqrt{2}}(0,v_2',0)^T~.\nn
\eea
The top vevs break \3311 to the SM gauge symmetry with $v_\sigma, w\gg v_{EW}$, while $v_{EW} = (v_1^2 +v_2^2 + v_2'^2)^{1/2} = 246$ GeV leads to electroweak breaking.  
Note that, while $w$ breaks $SU(3)_L\otimes U(1)_X$, $v_\sigma$ breaks $U(1)_N$.
When $\sigma$ and $\chi$ acquire similar vevs, $v_\sigma\sim w$, the two steps of the symmetry breaking process occur at the same time. 
\be 
SU(3)_c \otimes SU(3)_L \otimes U(1)_X \otimes U(1)_N \xrightarrow{v_\sigma, w} SU(3)_c \otimes SU(2)_L \otimes U(1)_Y \otimes M_P~.
\ee 

\begin{table}[t]
    \centering
    \begin{tabular}{|c|c|c|c|c|c|c|c|}
        \hline
\hspace{0.2cm} Field \hspace{0.2cm}& \hspace{0.2cm}SU(3)$_c$ \hspace{0.2cm}&\hspace{0.2cm} SU(3)$_L$ \hspace{0.2cm}& \hspace{0.2cm}U(1)$_X$ \hspace{0.2cm}& \hspace{0.2cm}U(1)$_N$  \hspace{0.2cm}&\hspace{0.2cm} $Q$\hspace{0.2cm} &\hspace{0.2cm} $B-L$\hspace{0.2cm} &\hspace{0.2cm} $M_P = (-1)^{3(B-L)+2s}$ \hspace{0.2cm}\\
\hline\hline
        $q_{i L}$ & {\bf 3} &$ \overline{{\mathbf 3}} $& 0 & 0 & $(-\frac{1}{3},\frac{2}{3},-\frac{1}{3})^T$&$(\frac{1}{3},\frac{1}{3},-\frac{2}{3})^T$&$(++-)^T$ \\
       $ q_{3L}$ & {\bf 3} & {\bf 3} & $\frac{1}{3}$ & $\frac{2}{3}$ & $(\frac{2}{3},-\frac{1}{3},\frac{2}{3})^T$& $(\frac{1}{3},\frac{1}{3},\frac{4}{3})^T$&$(++-)^T$ \\
        $u_{aR}$ & {\bf 3} & {\bf 1} & $\frac{2}{3}$ & $\frac{1}{3}$ &  $\frac{2}{3}$&  $\frac{1}{3}$&$+$ \\
        $d_{aR}$ & {\bf 3} & {\bf 1} & $-\frac{1}{3}$ & $\frac{1}{3}$ &  $-\frac{1}{3}$&  $\frac{1}{3}$&$+$ \\
        $U_{3R}$ & {\bf 3} & {\bf 1} & $\frac{2}{3}$ & $\frac{4}{3}$ &$\frac{2}{3}$&$\frac{4}{3}$&$-$ \\
        $D_{i R}$ & {\bf 3} & {\bf 1} & $-\frac{1}{3}$ & $-\frac{2}{3}$ & $-\frac{1}{3}$&$-\frac{2}{3}$&$-$ \\
        $l_{aL}$ & {\bf 1} & {\bf 3} & $-\frac{1}{3}$ & $-\frac{2}{3}$ &$(0,-1,0)^T$&$(-1,-1,0)^T$&$(++-)^T$ \\
       $ e_{a R}$ & {\bf 1} & {\bf 1} & $-1$ & $-1$ & $-1$&$-1$&$+$ \\
       \hline
       $ \nu_{i R}$ & {\bf 1} & {\bf 1} & $0$ & $-4$ & $0$& $-4$&$-$ \\
       $  \nu_{3R}$ & {\bf 1} & {\bf 1} & $0$ & $5$ & $0$& $5$&$+$ \\ $  N_{aR}$ & {\bf 1} & {\bf 1} & $0$ & $0$ & $0$& $0$& $-$  \\
       \hline\hline
        $\eta$ & {\bf 1} & {\bf 3} & $-\frac{1}{3}$ & $\frac{1}{3}$ &  $(0,-1,0)^T$&  $(0,0,1)^T$&$(++-)^T$ \\
        $\rho$ & {\bf 1} & {\bf 3} & $\frac{2}{3}$ & $\frac{1}{3}$ &  $(1,0,1)^T$&  $(0,0,1)^T$&$(++-)^T$ \\
        $\chi$ & {\bf 1} & {\bf 3} & $-\frac{1}{3}$ & $-\frac{2}{3}$ &  $(0,-1,0)^T$&  $(-1,-1,0)^T$&$(--+)^T$ \\
        $\sigma$ & {\bf 1} & {\bf 1} & $0$ & $2$ &  $0$ &  $2$&$+$ \\
       \hline
        $\zeta$ & {\bf 1} & {\bf 3} & $\frac{2}{3}$ & $\frac{7}{3}$ &  $(1,0,1)^T$&  $(2,2,3)^T$&$(+,+,-)^T$ \\
        $\xi$ & {\bf 1} & {\bf 3} & $\frac{2}{3}$ & $\frac{4}{3}$ &  $(1,0,1)^T$&  $(1,1,2)^T$&$(-,-,+)^T$ \\
\hline    
\end{tabular}
\caption{Particle content of the theory. Here $a=1,2,3$ and $i =1,2$ are family indices.}
 \label{tab}
\end{table}

The last step takes place when the first and second components of the triplets acquire vevs, and we are left with 
\be 
SU(3)_c \otimes SU(2)_L \otimes U(1)_Y \otimes M_P~\xrightarrow{v_1, v_2, v_2'}~SU(3)_c \otimes U(1)_Q \otimes M_P~,
\ee 
in such a way that $(v_1^2+v_2^2+v_2'^2)^{1/2} = v_{EW}$, the electroweak scale.

This process leaves at the end a matter-parity symmetry, $M_P$, defined as 
\bea\label{MP}
    M_P &=& (-1)^{3(B-L) + 2s}~.
\eea

Notice the important fact that only the $M_P$-even scalar fields get vevs. This implies that matter-parity remains as an absolutely conserved residual gauge symmetry 
even after spontaneous symmetry breaking, implying that the lightest amongst the $M_P$-odd particles is stable. 
Here we notice that the presence of the nonvanishing vev $v_\sigma$ breaks $U(1)_N$ at a potentially large scale, preventing the appearance of a light $Z'$ gauge boson. 


\section{Neutrino masses}  
\label{sec:neutrino-masses}

\begin{figure}[h!]
\centering
\includegraphics[scale=.5]{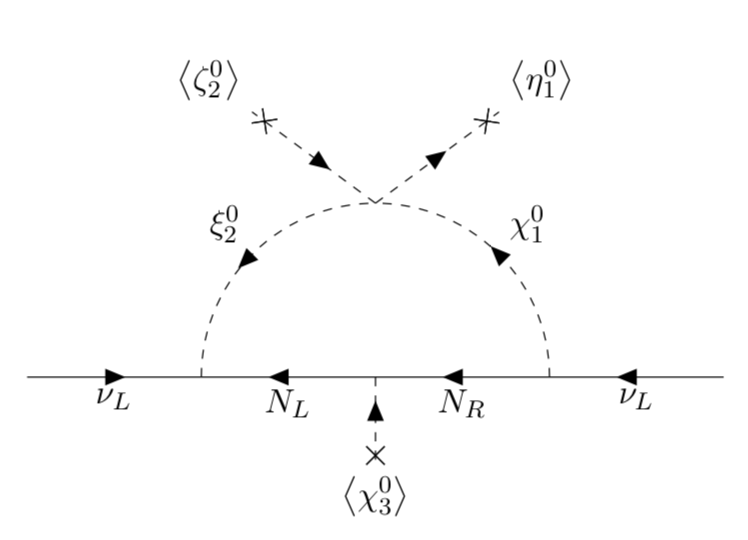}
\caption{\centering One-loop ``scotogenic'' neutrino mass.}
\label{fig1}
\end{figure}

Taking into account the leptons and scalars present in our model, as shown in Table \ref{tab}, the following Yukawa sector can be written down 
\bea\label{Yuk}
- \mathcal{L}_{lep} &=& y^e_{ab}\,\overline{l_{aL}}\, \rho\, e_{bR} + y^N_{ab}\, \overline{l_{aL}}\, \chi N_{bR} + h_{ab}\, \overline{l_{aL}}\,(l_{bL})^c\, \xi^* +   \frac{(m_N)_{ab}}{2}\, \overline{(N_{aR})^c}\,N_{bR} + h.c.~, 
\eea
where $y^e, y^N, h$ and $m_N$ are complex $3\times3$ matrices, where $m_N$ is symmetric, due to the Pauli principle. In contrast, due again to the symmetry structure of the theory, the Yukawa coupling matrix $h$ is anti-symmetric in family space.
Notice that this anti-symmetric Yukawa coupling was first proposed in~\cite{Valle:1983dk}. 
While the original scheme is no longer viable, given the current neutrino oscillation data, the new construction provides a consistent variant that also 
accounts for WIMP dark matter in a scotogenic way, i.e. dark matter emerges as a neutrino mass mediator, see Fig.~\ref{fig1}.

Notice that, while the fields $\nu_{iR}$ and $\nu_{3R}$ with
non-standard charges~\cite{Ma:2014qra,Ma:2015raa,Ma:2015mjd} are necessary in order to ensure anomaly
cancellation, such choice of charges forbids their coupling to the
other leptons as well as scalars, justifying their absence from the
Lagrangian given above~\footnote{$\nu_{iR}$ and $\nu_{3R}$ masses
  could be generated, for example, by coupling them to new scalars
  transforming as $({\bf1},{\bf1}, 0,8)$ and $({\bf1},{\bf1}, 0,-10)$,
  respectively. Note that matter-parity conservation would not be
  spoiled when such scalars acquire vevs.}. 

The first term in Eq. (\ref{Yuk}) generates a mass term to the charged leptons when $\rho$ acquires a vev:  
\be 
M^e = y^e\frac{v_2}{\sqrt{2}}~,
\ee 
where the family indices have been omitted.
The neutral leptons $N_{iL}$ and $N_{iR}$ mass matrix is given as 
\be 
\label{eq:Nmass}
M^N = \frac{1}{2}\begin{pmatrix}
 0 & y^N w \\
 (y^N)^T w & m_N~
\end{pmatrix}
\ee 
in the basis $(N_L, (N_R)^c)^T$. Such a matrix is diagonalised (in the one family approximation) as
\be \label{eq:Nmix}
\begin{pmatrix} N_1\\N_2\end{pmatrix}=\begin{pmatrix} \cos{\theta_N} & -\sin{\theta_N} \\
\sin{\theta_N} & \cos{\theta_N} \end{pmatrix}
\begin{pmatrix}
N_L \\(N_R)^c 
\end{pmatrix}
 \quad \mbox{with}\quad\tan(2\theta_N)=\frac{2y^N w}{m_N}.
\ee

Turning to the light neutrinos $\nu_L$, it is easy to see that tree-level mass terms could be generated if the first component of
$\chi$ or the second of $\xi$ acquired a vev. This, however, does not occur as a result of the assumed pattern of vevs and this, in turn, is 
dynamically consistent with the minimization of the potential.
This way matter-parity conservation emerges as a residual symmetry. 

Neutrino masses are radiatively generated by the one-loop diagram in Fig.\ref{fig1}. The relevant scalar interaction is
the one governed by the $\lambda_1$, see Eq.(\ref{V}), and leads to
\begin{align}
\label{eq:mnu}
m_{\nu}^{ab}&=\frac{1}{8\pi^2} h^{* ac} s_N c_N c_1 \left\{m_{N_1}\left[s_{S_2} c_{S_2}\left(Z\left(\frac{m_{S_1}^2}{m_{N_1}^2}\right)-Z\left(\frac{m_{S_2}^2}{m_{N_1}^2}\right)\right)-s_{A_2}c_{A_2}\left(Z\left(\frac{m_{A_1}^2}{m_{N_1}^2}\right)-Z\left(\frac{m_{A_2}^2}{m_{N_1}^2}\right)\right)\right]\right. \\
&\left.- m_{N_2}\left[s_{S_2} c_{S_2}\left(Z\left(\frac{m_{S_1}^2}{m_{N_2}^2}\right)-Z\left(\frac{m_{S_2}^2}{m_{N_2}^2}\right)\right)-s_{A_2} c_{A_2}\left(Z\left(\frac{m_{A_1}^2}{m_{N_2}^2}\right)-Z\left(\frac{m_{A_2}^2}{m_{N_2}^2}\right)\right)\right]\right\}_{cd} y^{N* db} + \left\{a\leftrightarrow b\right\}, \nn
\end{align}
where $s_x\equiv \sin\theta_x,c_x\equiv \cos\theta_x$, and the loop function $Z(x)$ is defined as 
\begin{align}
\label{eq:zfunc}
Z(x)&=\frac{x}{1-x}\text{ln}x.
\end{align}
An important point to note here is that owing to the antisymmetry of the Yukawa matrix $h$, the resulting neutrino mass matrix of Eq. \eqref{eq:mnu} is of rank two.
This implies that when rotated to the mass basis, only two neutrinos acquire mass and one remains massless.
This unique feature provides a novel origin for the masslessness of one neutrino~\footnote{This is reminiscent of the proposal in~\cite{Valle:1983dk}, currently ruled out by the oscillation data.}
that should be contrasted with the usual models for one massless neutrino, which typically rely on missing partner mechanisms.

Note that the matter-parity odd neutral fermions $\left(N_L,(N_R)^{c}\right)$ are obtained from Eq.~(\ref{eq:Nmix}),
while the scalar masses $m_{1,2}$ are given in Eq.~(\ref{eq:s_mass}) and the mixing angles of $\left(\xi_2^0, \chi_1^0, \eta_3^0\right)_{S,A}$ 
come from Eq.~(\ref{eq:s_mpodd_mix}). 

It is worth pointing out that, in addition to the usual loop suppression characteristic of scotogenic models, our result in Eq. (\ref{eq:mnu}) is further suppressed by the factor $c_1\sim v_1/w\ll 1$, see Eq. (\ref{theta1}). This is needed in order to identify the physical mass eigenstates associated with the scalar mediators in the scotogenic loop.

All fields running inside the neutrino mass loop are odd under
matter-parity. The exact conservation of this symmetry implies that
the lightest among the $M_P$-odd particles is stable, and therefore can
play the role of dark matter. Thus, the present model generates
``scotogenic'' neutrino masses, with the crucial dark matter
stabilising symmetry emerging naturally as a residual subgroup of the
original gauge symmetry.


\section{Scalar sector}
\label{sec:scalar-sector}


In addition to the three SU(3)$_L$ triplets  $\eta,\, \chi,\, \rho$ our model employs two others,  $\xi$ and $\zeta$.
The scalar triplets can be decomposed into
\bea 
\eta, \chi \equiv \begin{pmatrix} \eta_1^0 \\ \eta_2^- \\ \eta_3^0 \end{pmatrix}~,\begin{pmatrix} \chi_1^0 \\ \chi_2^- \\ \chi_3^0 \end{pmatrix}\quad \quad  \rho, \xi, \zeta \equiv \begin{pmatrix} \rho_1^+ \\ \rho_2^0 \\ \rho_3^+ \end{pmatrix}~,\begin{pmatrix} \xi_1^+ \\ \xi_2^0 \\ \xi_3^+ \end{pmatrix}~,\begin{pmatrix} \zeta_1^+ \\ \zeta_2^0 \\ \zeta_3^+ \end{pmatrix}~,
\eea
and the neutral components, as well as the scalar singlet $\sigma$, can be further decomposed into their CP-even (S) and
CP-odd (A) parts, in such a way that for a given neutral scalar field $s_i^0$, we have 
\be 
s_i^0 \equiv \frac{1}{\sqrt{2}} (v_{s_i} + S_{s_i} + i A_{s_i})~,
\ee 
with $s$ denoting generically all the scalars, and 
$v_{\eta_1} = v_1, v_{\rho_2} = v_2, v_{\chi_3} = w, v_{\zeta_2} = v_2' $, and $v_\sigma = v_\sigma$, 
whereas all the other vevs vanish, as already discussed in Eq. (\ref{vevs}). 

Given the five scalar triplets and the singlet in Table \ref{tab}, the scalar potential can be written as 
\bea\label{V}
V &=& \sum_{s} \left[\mu_s^2 (s^\dagger s)+\frac{\lambda_s}{2} (s^\dagger s)^2   \right] + \sum_{s_1,s_2}^{s_1>s_2} \left[ \lambda_{s_1 s_2 }(s_1^\dagger s_1)(s_2^\dagger s_2) \right] + \sum_{t_1,t_2}^{t_1>t_2} \left[ \lambda'_{t_1 t_2  }(t_1^\dagger t_2)(t_2^\dagger t_1)\right]\\
&& + \frac{\mu_1}{\sqrt{2}} \eta\rho\chi  + \frac{\mu_2}{\sqrt{2}} (\zeta^\dagger\rho)\sigma  + \lambda_1(\chi^\dagger \eta)(\zeta^\dagger \xi) +  \lambda_2  (\chi^\dagger \xi)(\zeta^\dagger \eta)  + \lambda_3 (\chi^\dagger \eta)(\xi^\dagger \rho)\nn\\
&& + \lambda_4 (\chi^\dagger \rho)(\xi^\dagger \eta)
+ \lambda_5(\eta\zeta\chi)\sigma^* + h.c. ~,\nn
\eea 
where $t$ only varies through all the scalar triplets: $t = \eta,\, \chi,\, \rho,\,\xi,\,\zeta$, while $s$ varies through all the scalars, {\it i.e.} the triplets in $t$ plus the singlet $\sigma$. 

By minimising the scalar potential, we obtain the ``tadpole'' conditions
\bea 
\mu_1 v_2 w+\lambda_5v_2'w v_\sigma + v_1 \left(2 \mu_\eta^2 +\lambda_{\eta}v_1^2 + \lambda_{\eta \rho} v_2^2 + \lambda_{\eta \zeta}  v_2'^2 + \lambda_{\eta \chi} w^2+\lambda_{\eta\sigma}v_\sigma^2 \right) &=& 0~,\\
\mu_1 v_1 w + \mu_2 v_2' v_\sigma + v_2 \left[ 2 \mu_\rho^2 + \lambda_{\eta \rho}v_1^2 +\lambda_\rho v_2^2 + (\lambda_{\rho \zeta} + \lambda_{\rho \zeta 2} )v_2'^2  + \lambda_{\rho \chi}w^2 +\lambda_{\rho\sigma}v_\sigma^2\right] &=& 0~,\nn\\
\mu_1 v_1 v_2 +\lambda_5 v_1 v_2' v_\sigma+ w \left(2 \mu_\chi^2+\lambda_{\eta \chi}v_1^2 +\lambda_{\rho \chi} v_2^2 + \lambda_{\chi \zeta} v_2'^2 + \lambda_\chi   w^2+\lambda_{\chi\sigma}v_\sigma^2\right) &=& 0~,\nn\\
 \mu_2 v_2 v_\sigma  + \lambda_5 v_1 v_\sigma w+v_2' \left[2 \mu_\zeta^2 + \lambda_{\eta \zeta} v_1^2 + (\lambda_{\rho \zeta} + \lambda_{\rho \zeta 2}) v_2^2  + \lambda_\zeta v_2'^2 + \lambda_{\chi \zeta}  w^2 + \lambda_{\zeta\sigma} v_\sigma^2\right] &=& 0~,\nn\\
 \mu_2 v_2 v_2'+ \lambda_5 v_1 v_2' w+v_\sigma\left(2\mu_\sigma^2+\lambda_{\eta\sigma}v_1^2+\lambda_{\rho\sigma}v_2^2+ \lambda_{\zeta\sigma} v_2'^2+\lambda_{\chi\sigma} w^2 +\lambda_\sigma v_\sigma^2\right) &=&0~, \nn
\eea 
through which $\mu_\eta, \mu_\rho, \mu_\chi, \mu_\zeta$ and
$\mu_\sigma$ can be eliminated from the potential.
Nine out of the initial degrees of freedom in the scalar sector are
absorbed as longitudinal components of the massive gauge vector
bosons, $Z, Z', Z'', U^0, (U^0)^\dagger, W^\pm, V^{\pm}$. The
remaining scalar fields become massive, as we now discuss.

First, we focus on the scalar fields that enter the neutrino mass loop,
for which we show the corresponding mass matrices and diagonalise them
in Sec.~\ref{sec:neutr-mass-medi}, providing the mass eigenvalues and
eigenstates. For the other scalars, the mass matrices are given in
Sec.~\ref{App}.


\subsection{Neutrino-mass-mediator scalars}
\label{sec:neutr-mass-medi}


The scalar fields relevant to the neutrino mass loop in Fig. \ref{fig1} are part of the set of 
the $M_P$-odd neutral fields and can be grouped together into a CP-even and a CP-odd set:
$(S_{\xi_2},S_{\chi _1}, S_{\eta_3})$ and $(A_{\xi_2},A_{\chi _1}, A_{\eta_3})$, respectively. In such bases, we
can write down the following squared mass matrices
\bea\label{MSA}
M_{S,A}^2 &=&
\frac{1}{2}\begin{pmatrix}
a_{11} &  a_{12} & a_{13} \\ a_{12}  & a_{22} & a_{23}  \\
  a_{13} & a_{23} & a_{33}
\end{pmatrix}_{S,A}~,
\eea
where the elements $a_{ij}$ are defined as 
\bea 
(a_{11})_{S,A} &=& a_{11} = \lambda_{\eta \xi}  v_1^2 +  (\lambda_{ \rho \xi} + \lambda_{\rho \xi 2}) v_2^2+(\lambda_{  \xi\zeta} + \lambda_{\xi\zeta 2}) v_2'^2 + \lambda_{\chi \xi} w^2 +\lambda_{\xi\sigma} v_\sigma^2+ 2 \mu_\xi^2~,\\
(a_{22})_{S,A} &=& a_{22} =  \lambda_{\eta \chi 2}v_1^2-\frac{v_1}{w}(\lambda_5 v_2' v_\sigma + \mu_1  v_2 )  ~,\nn\\
(a_{33})_{S,A} &=& a_{33} =  \lambda_{\eta \chi 2}w^2-\frac{w}{v_1}(\lambda_5 v_2' v_\sigma + \mu_1  v_2 ) ~,\nn\\
(a_{12})_{S,A} &=& v_1(\lambda_1 v_2'\pm \lambda_3 v_2)~,\nn\\
(a_{13})_{S,A} &=& w(\lambda_3 v_2 \pm\lambda_1 v_2')~,\nn\\
(a_{23})_{S,A} &=& \pm(\lambda_{\eta \chi 2}v_1 w -\mu_1 v_2-\lambda_5 v_2' v_\sigma)~.\nn
\eea 

Each matrix has a vanishing eigenvalue associated with a would-be Goldstone boson that is absorbed by the gauge sector, more
specifically by the complex neutral gauge field $U_0$. We can find the massless eigenstate by rotating the second and third 
components of both the CP-even and CP-odd basis by
\be\label{theta1}
(\theta_{1})_{S,A} = \pm\arctan\left(\frac{ w}{v_1}\right)~,
\ee
respectively. After these transformations, the matrices in Eq. (\ref{MSA}) become 
\be 
\tilde{M}_{S,A}^2= \frac{1}{2}\begin{pmatrix}
 a_{11} &  x\, a_{12} & 0 \\
 x\,a_{12} &  x^2\,a_{22} & 0\\
 0 & 0 & 0
\end{pmatrix}_{S,A}~,\quad\mbox{with}\quad x^2 =\frac{v_1^2+w^2}{v_1^2}~. 
\ee 
Such matrices can be finally diagonalised by rotating the two first
components of each basis by
\be\label{theta2}
(\theta_{2})_{S,A} = \frac{1}{2}\arctan\left[\frac{2
      x\,(a_{12})_{S,A}}{a_{11}-x^2\,a_{22}}\right]~,
\ee
respectively.  By doing so, we obtain the eigenvalues
\bea
\label{eq:s_mass}
(m_{1,2}^2)_{S,A} &=& \frac{1}{2}\left[a_{11}+ x^2 a_{22}\pm\sqrt{(a_{11}-x^2 a_{22})^2+ 4 x^2 (a_{12})^2_{S,A}}\right] ~,\quad (m_{3}^2)_{S,A} = 0.
\eea 
In summary, the mass and flavour states can be related as
\be 
\label{eq:s_mpodd_mix}
 \begin{pmatrix}
 (S,A)_{m_1} \\
 (S,A)_{m_2} \\
 (S,A)_{m_3} \\
\end{pmatrix} =  \begin{pmatrix}
 \cos{\theta_2} & \sin{\theta_2} & 0\\
 -\sin{\theta_2} & \cos{\theta_2} & 0 \\
 0 & 0 & 1
\end{pmatrix}_{S,A}
 \begin{pmatrix}
 1 & 0 & 0\\
 0 & \cos{\theta_1} & \sin{\theta_1} \\
 0 & -\sin{\theta_1} & \cos{\theta_1}
\end{pmatrix}_{S,A}
\begin{pmatrix}
 (S,A)_{\xi_2} \\
 (S,A)_{\chi_1} \\
 (S,A)_{\eta_3} \\
\end{pmatrix}~.
\ee 


\subsection{Mass matrices of the other scalars} 
\label{App}


In this section we present the squared mass matrices associated with
the scalar fields that do not take part in the neutrino mass loop.
The CP-even and $M_P$-even neutral fields can be grouped in the basis
$(S_{\eta_1}, S_{\rho_2}, S_{\chi_3}, S_{\zeta_2}, S_\sigma)$, so that we have
the following symmetric squared mass matrix 
\be \label{MS2}
M_{S_2}^2 =
\frac{1}{2}
\begin{pmatrix}
b_{11} & 2\lambda_{ \eta \rho}v_1 v_2  +  \mu_1 w  & 2\lambda_{\eta \chi}v_1 w  + \mu_1 v_2+\lambda_5 v_2' v_\phi &  2\lambda_{\eta \zeta} v_1 v_2'+\lambda_5 v_\sigma w &  2\lambda_{\eta \sigma} v_1 v_\sigma+\lambda_5 v_2' w   \\
 \star  & b_{22} & 2\lambda_{\rho \chi}v_2 w  + \mu_1 v_1  &  2(\lambda_{\rho \zeta} + \lambda_{\rho \zeta 2} ) v_2 v_2'  +\mu_2 v_\sigma & 2 \lambda_{\rho\sigma} v_2 v_\sigma + \mu_2 v_2'\\
 \star & \star & b_{33} &  2\lambda_{\chi \zeta} v_2' w +\lambda_5 v_1 v_\sigma  &  2\lambda_{\chi \sigma} v_\sigma w +\lambda_5 v_1 v_2' \\
 \star & \star  &  \star & b_{44} & 2\lambda_{\zeta\sigma} v_2' v_\sigma+\mu_2 v_2+\lambda_5 v_1 w \\
 \star & \star  &  \star & \star &  b_{55}\\
\end{pmatrix}~,
\ee
with the diagonal elements given by
\bea
b_{11} &=&
2\lambda_\eta v_1^2  -\frac{w}{ v_1}\left(\mu_1 v_2+\lambda_5 v_2' v_\sigma\right)~, \\
b_{22} &=& 2\lambda_\rho v_2^2  -\frac{ \mu_1 v_1 w + \mu_2 v_2' v_\sigma }{v_2}~,\nn\\
b_{33} &=& 2 \lambda_\chi w^2 -\frac{v_1}{w}\left(\mu_1 v_2 +\lambda_5 v_2' v_\sigma\right)~,\nn\\
b_{44} &=& 2\lambda_\zeta v_2'^2-\frac{v_\sigma}{v_2'}\left(\mu_2 v_2+\lambda_5 v_1 w \right)~,\nn\\
b_{55} &=&2\lambda_\sigma v_\sigma^2 -\frac{v_2'}{v_\sigma}\left(\mu_2 v_2 + \lambda_5 v_1 w \right)~.\nn
\eea 
Upon diagonalisation, five non-vanishing masses appear associated with five physical scalars, one of which is the $125$ GeV Higgs boson discovered at the LHC. \\[-.2cm]

Taking into account now the CP-odd, $M_P$-even fields we obtain the squared mass matrix below, expressed in the basis
$ (A_{\eta_1},A_{\rho_2},A_{\chi_3}, A_{\zeta_2},A_\sigma)$, 
\be\label{MA2}
M_{A_2}^2 = 
\frac{1}{2}\begin{pmatrix}
 -\frac{w}{v_1}( \mu_1 v_2+ \lambda_5 v_2' v_\sigma)  & -  \mu_1 w  & -\mu_1 v_2 - \lambda_5 v_2' v_\sigma & -\lambda_5 v_\sigma w & \lambda_5 v_2' w \\
\star  & -\frac{\mu_1 v_1 w+\mu_2 v_2' v_\sigma}{v_2} & -\mu_1 v_1  & \mu_2 v_\sigma & -\mu_2 v_2'\\
 \star  &\star  & -\frac{v_1 }{w}(\mu_1 v_2+\lambda_5 v_2' v_\sigma ) & -\lambda_5 v_1 v_\sigma & \lambda_5 v_1 v_2'  \\
 \star  & \star  & \star  & -\frac{v_\sigma}{v_2'}(\mu_2 v_2+ \lambda_5v_1 w) & \mu_2 v_2 +\lambda_5 v_1 w \\
 \star  & \star  & \star  &  \star  & -\frac{v_2'}{v_\sigma}(\mu_2 v_2 + \lambda_5 v_1 w) 
\end{pmatrix}~.
\ee
Three states remain massless and are absorbed by the neutral gauge
bosons $Z, Z^\prime, Z^{\prime\prime}$.  The other two states give rise to two massive CP-odd scalars. \\[-.2cm]

At last, we consider the charged scalar fields. 
In the $M_P$-even basis $(\eta^\pm_2,\rho_1^\pm, \zeta_1^\pm,\xi_3^\pm)$, we can write the first squared mass matrix as
\be\label{CSMM1}
(M_{1}^\pm)^2 =
\frac{1}{2}\begin{pmatrix} c_{11} &  \lambda_{\eta \rho 2} v_1 v_2- \mu_1 w & \lambda_{\eta \zeta 2} v_1 v_2' - \lambda_5 v_\sigma w & \lambda_2 v_2' w \\
 \lambda_{\eta \rho 2}v_1 v_2 -  \mu _1 w &  c_{22} & \lambda_{\rho \zeta 2}v_2 v_2' + \mu_2 v_\sigma & 0 \\
 \lambda_{\eta \zeta 2} v_1 v_2'-\lambda_5 v_\sigma w  &  \lambda_{\rho \zeta 2} v_2 v_2' + \mu_2 v_\sigma & c_{33} & \lambda_2 v_1   w \\
\lambda_2 v_2' w & 0 &  \lambda_2 v_1 w & c_{44} \\
\end{pmatrix}
\ee
with 
\bea 
c_{11} &=& 
 \lambda_{\eta \rho 2} v_2^2 +  \lambda_{\eta \zeta 2} v_2'^2 -\frac{ w }{v_1}(\mu_1  v_2+\lambda_5 v_2' v_\sigma)~,\\
c_{22} &=&  \lambda_{\eta \rho 2} v_1^2 - \lambda_{\rho \zeta 2}v_2'^2 -\frac{ \mu_1 v_1 w+ \mu_2 v_2' v_\sigma}{v_2}~,\nn\\
c_{33} &=&  \lambda_{\eta \zeta 2} v_1^2- \lambda_{\rho \zeta 2}v_2^2-\frac{\mu_2 v_2 v_\sigma+\lambda_5 v_1 v_\sigma w}{v_2'}~,\nn\\
c_{44}&=& \lambda_{ \eta \xi}  v_1^2 +  \lambda_{\rho \xi} v_2^2 +  \lambda_{\xi\zeta} v_2'^2 +  (\lambda_{\chi \xi} +\lambda_{\chi \xi 2})w^2 +\lambda_{\xi\sigma}v_\sigma^2+2 \mu_\xi^2.\nn
\eea 
Whereas in the $M_P$-odd basis $(\xi_1^\pm, \chi_2^\pm, \rho_ 3^\pm, \zeta_3^\pm)$, we write down
\be \label{CSMM2}
(M_2^\pm)^2=
\frac{1}{2}
\begin{pmatrix}
 d_{11} & \lambda_4 v_1 v_2 & \lambda_4 v_1 w & 0 \\
   \lambda_4 v_1 v_2 &  d_{22} & \lambda_{\rho \chi 2} v_2 w - \mu_1 v_1 & \lambda_{\chi \zeta 2} v_2'  w -\lambda_5 v_1 v_\sigma\\
  \lambda_4 v_1 w & \lambda_{\rho \chi 2} v_2 w -  \mu_1 v_1  & d_{33} & \lambda_{\rho \zeta 2} v_2 v_2' + \mu_2 v_\sigma   \\
 0 & \lambda_{\chi \zeta 2} v_2'  w - \lambda_5 v_1 v_\sigma  &  \lambda_{\rho \zeta 2} v_2 v_2' +\mu_2 v_\sigma & d_{44} \\
\end{pmatrix}
\ee 
with 
\bea 
d_{11} &=& (\lambda_{\eta \xi} +\lambda_{\eta \xi 2}) v_1^2+ \lambda_{\rho \xi}v_2^2+ \lambda_{\xi \zeta}v_2'^2+ \lambda_{\chi \xi}  w^2+ \lambda_{\xi\sigma} v_\sigma^2 + 2 \mu_\xi^2~,\\
d_{22} &=& \lambda_{\rho \chi 2} v_2^2  + \lambda_{\chi \zeta 2} v_2'^2 - \frac{ v_1}{w}( \mu_1  v_2 + \lambda_5 v_2' v_\sigma) ~,\nn\\
d_{33} &=&  \lambda_{\rho \chi 2}w^2- \lambda_{\rho \zeta 2} v_2'^2-\frac{\mu_1 v_1 w + \mu_2 v_2' v_\sigma}{v_2}  ~,\nn\\
d_{44} &=&  \lambda_{\chi \zeta 2} w^2 - \lambda_{\rho \zeta 2} v_2^2 -\frac{v_\sigma}{v_2'}(\mu_2 v_2  + \lambda_5 v_1  w) ~.\nn
\eea 
Each of the squared mass matrices above has a vanishing eigenvalue associated with a would-be Goldstone boson that will be absorbed by the charged gauge bosons $W^{\pm}$ and $V^\pm$. Finally, we are left with six heavy charged scalar fields in the model.


\section{scotogenic dark matter neutrinoless double beta decay}
\label{sec:0nbb}


This model can harbour a WIMP dark matter candidate that can be either scalar or a fermion.

Whatever the dark matter profile will be, we note the presence of new interactions, in addition to the standard processes of the simplest scotogenic model, 
which are primarily responsible for setting the dark matter relic abundance.
In particular there are new t-channel processes involving the matter-parity-odd electrically neutral gauge boson connecting the same-charge
components of the fermion triplets.

Concerning dark matter detection, let us recall that in the simplest scotogenic scenario~\cite{Ma:2006km}, 
it proceeds primarily through the Higgs portal.
In our model this portal has additional contributions, thanks to the presence of new scalar bosons.
Moreover, since the dark sector particles carry $B-L$ charges, the usual Higgs portal is accompanied by a $Z'$ portal.
Which of the two portals will be dominant depends on the $B-L$ breaking scale, as well as on the various coupling strengths,
particularly the Higgs-dark matter quartic coupling and $B-L$ gauge coupling strength $g'$.  
In the limit of large $v_\sigma$ and $w$ we recover the standard scotogenic dark matter phenomenology, which has been investigated before in~\cite{Kang:2019sab}.
In contrast, if the $B-L$ breaking scale is low the $Z'$ may become significant. However, both scenarios have been well studied in the
literature see, e.g.~\cite{Ma:2015mjd} for the $Z'$ portal, so they will not be analysed here.
Instead, we move directly to neutrinoless double beta decay, which presents interesting characteristic features. \\

\begin{figure}[ht!]
\centering
\includegraphics[scale=.4]{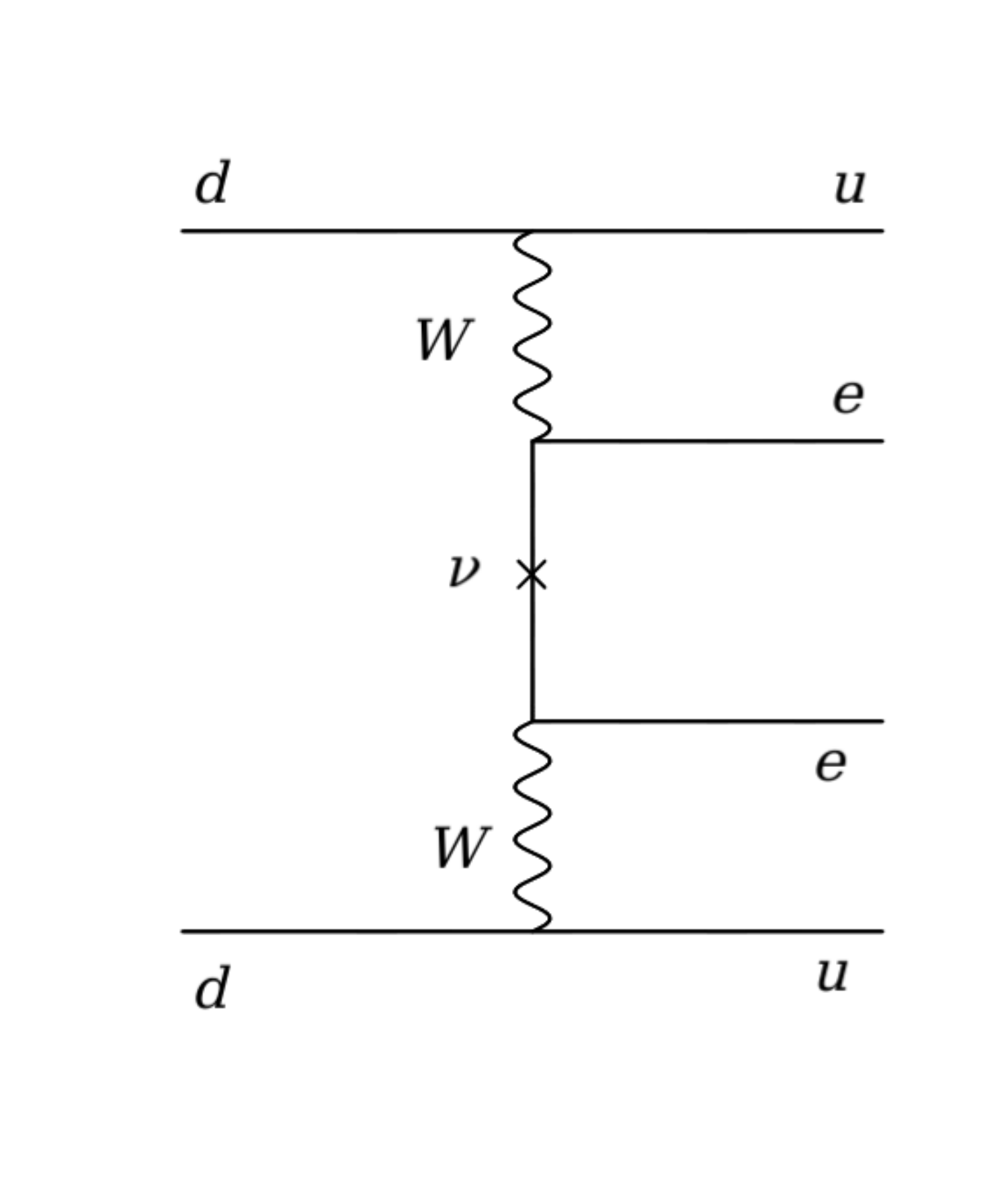}
\caption{\centering Standard ``mass-mechanism'' $0\nu\beta\beta$ contribution}
\label{fig2}
\end{figure}

Whether neutrinos are Majorana or Dirac fermions is still an open question. The case for Majorana neutrinos, as predicted by our model, can be undoubtedly established if neutrinoless double beta decay is ever observed \cite{Schechter:1981bd}.
The standard light neutrino-mediated $0\nu\beta\beta$ decay contribution is shown in Fig \ref{fig2}.
Its amplitude involves the lightest charged gauge boson $W^\pm$ exchange and hence is expressed in terms of the Fermi 
constant $G_F$, the typical momentum exchange $p$ characterizing the process, and the effective Majorana mass $\vev{ m_{\beta\beta}}$ 
\be\label{mbb}
\vev{ m_{\beta\beta} } = |\cos^2 \theta_{12}\cos^2 \theta_{13} m_1+\sin^2 \theta_{12}\cos^2 \theta_{13} m_2 e^{2 i \phi_{12}} + \sin^2 \theta_{13} m_3 e^{2 i \phi_{13}}|~,
\ee 
is neatly expressed in the symmetric parametrization of the lepton mixing matrix~\cite{Schechter:1980gr} in terms of the mixing angles $\theta_{12}$ and $\theta_{13}$,
the physical Majorana phases~\cite{Schechter:1980gk} $\phi_{12}$ and $\phi_{13}$, and the neutrino mass eigenvalues $m_a$ obtained from Fig.~\ref{fig1}.

\begin{figure}[h!]
\centering
\includegraphics[scale=.5]{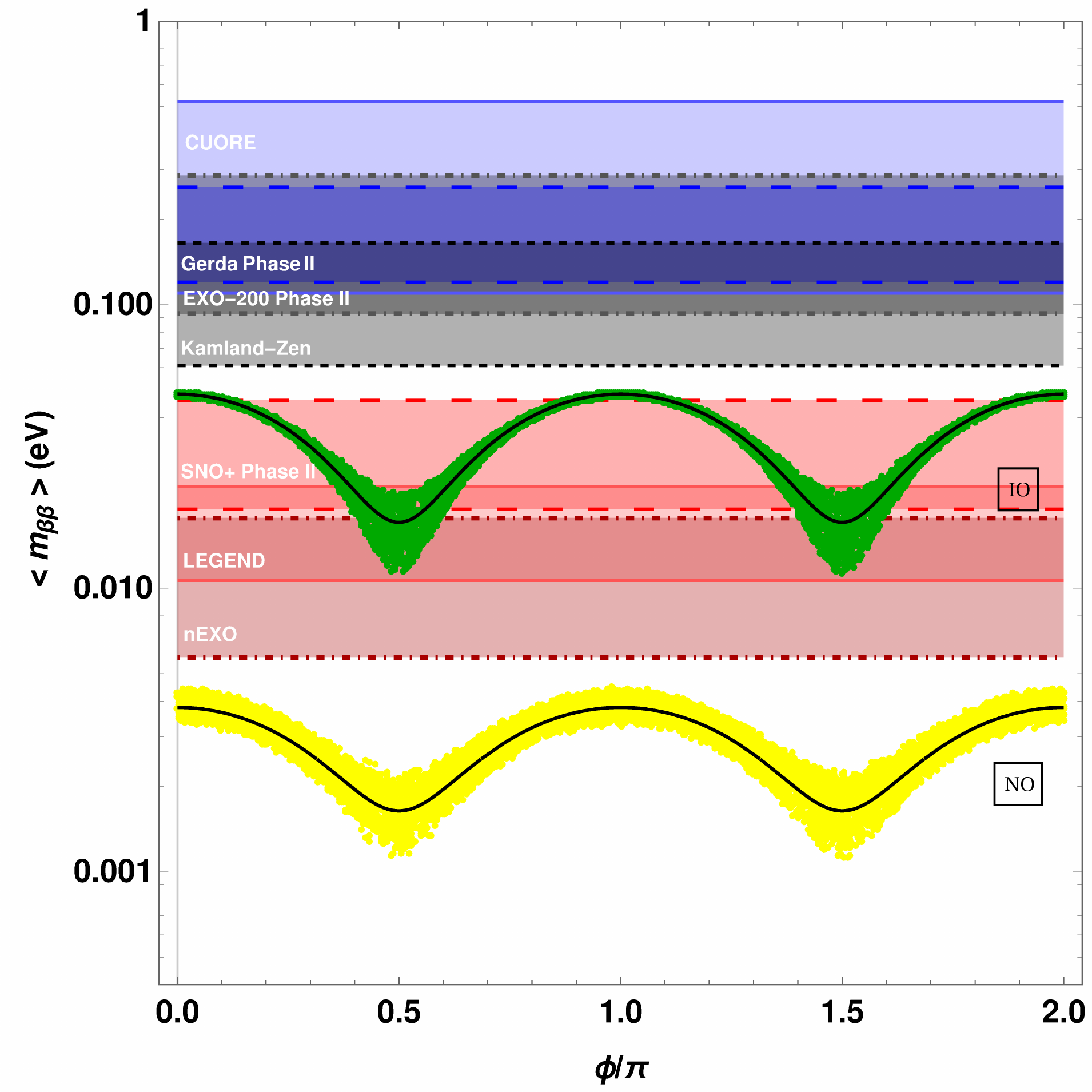} 
\caption{Effective Majorana mass vs relative Majorana phase. The lightest neutrino is massless because of
the anti-symmetry of the Yukawa coupling matrix $h_{ab}$ determining the loop-induced neutrino mass through Eq.~(\ref{eq:mnu}).
We give the expected $3\sigma$ \znbb bands for the case of inverted and normal mass ordering, in green and yellow, respectively. 
Horizontal bands represent current experimental limits and future sensitivities. }
\label{fig3}
\end{figure}

It is well-known that, in a generic model, this amplitude can vanish as a result of destructive interference amongst the three light neutrinos.
This actually can happen for normal-ordered neutrinos, currently preferred by oscillations~\cite{deSalas:2017kay}.
An important feature that emerges from the structure of our model is that one of the light neutrinos is predicted to be massless. 
In this case, with a massless neutrino in the spectrum, $\vev{ m_{\beta\beta} }$ is given in terms of just one free parameter, the relative Majorana phase: $\phi \equiv \phi_{12}-\phi_{13}$, all other parameters are fairly well-determined by the oscillation experiments.
One can easily verify that in this case the effective Majorana mass $\vev{ m_{\beta\beta}}$ never vanishes, even for the case of normal mass ordering, as shown in Fig. \ref{fig3}. 
Thus, thanks to the presence of a massless neutrino, our model is testable, at least for the inverted ordering (IO) case,  which falls within the expected sensitivity of the upcoming next generation $0\nu\beta\beta$ decay experiments.

The top four horizontal bands in Fig. \ref{fig3} represent the current experimental limits coming from CUORE ($\vev{m_{\beta \beta}} < 110 - 520$ meV)~\cite{Alduino:2017ehq},
EXO 200 Phase II ($93 - 286$ meV)~\cite{Anton:2019wmi}, Gerda Phase II ($120 - 260$ meV)~\cite{Agostini:2018tnm} and Kamland Zen ($61 - 165$ meV)~\cite{KamLAND-Zen:2016pfg}.
The widths of these bands reflect uncertainties in nuclear matrix elements.
The lower bands show the future sensitives from LEGEND ($10.7 - 22.8$ meV)~\cite{Abgrall:2017syy}, SNO + Phase II ($19 - 46$ meV)~\cite{Andringa:2015tza} and nEXO ($5.7 - 17.7$ meV)~\cite{Albert:2017hjq}.

Note that the prediction of a lower bound for the \znbb decay rate has been shown to occur in ``missing partner'' neutrino
mass models, such as the ``incomplete'' (3,2) seesaw mechanism containing only two isosinglet neutrinos~\cite{Schechter:1980gr}, or similar radiative mechanisms~\cite{Reig:2018ztc}.
However here it appears in a novel way, associated with the anti-symmetry of the Yukawa coupling matrix $h_{ab}$ determining the loop-induced neutrino mass through Eq.~(\ref{eq:mnu}).

Note that other tree-level contributions mediated by charged scalars are neglected,
since they are suppressed by the SU(3) symmetry breaking scale.

\section{Flavour changing neutral currents and colliders}
\label{sec:pheno}

In this section, we comment briefly on other important phenomenological features of the present model beyond scotogenic dark matter and neutrino masses. 
We consider possible new contributions to well-known processes, such as meson-anti-meson mass differences, as well as genuinely new processes 
where the new particles present in the model could be directly produced at particle colliders such as the LHC.

In the present model flavour changing neutral current (FCNC) processes can be induced at the tree-level.
Indeed, it is well-known that tree-level FCNCs appear in 331 models due to the embedding of quark families in different $SU(3)_L$ representations, 
as required by anomaly cancellation~\cite{Singer:1980sw}.
These FCNCs will be mediated by the new $Z^\prime$ gauge boson.
In the context of these models, there have been studies devoted to such FCNCs. 
One finds that, in order to be in agreement with experimental results, for example $B-\bar{B}$ mass difference, 
the new vector bosons must be heavier than a few TeV, see Fig.~5 of Ref.~\cite{Queiroz:2016gif}. 
Although in our extended 331 model, not one but two new vector bosons, $Z^\prime_{1,2}$, appear~\cite{Dong:2014wsa}, and both mediate FCNCs 
at tree-level~\footnote{There is also an unmixed electrically neutral gauge boson coupling neutrinos to a dark fermion.}, we expect the 
restrictions on the intermediate vector boson masses to be similar.
 
Another possible origin of tree-level FCNCs is through the mixing between standard and non-standard fermions. 
In the present model, however, thanks to matter-parity conservation, the latter does not take place, since standard 
and non-standard fermions have opposite matter-parities (see Table \ref{tab}) and, as such, do not mix.
Therefore, matter parity conservation plays yet another role: it forbids the mixing between ordinary and new fermions, 
thereby preventing the appearance of potentially dangerous FCNCs associated with it.

Scalar bosons can also mediate tree-level FCNCs. 
In order to find the terms that lead to flavour-changing currents, we focus on the quark sector and write down the corresponding Yukawa terms
\bea\label{qYuk}
-\mathcal{L}_q &=& y^d_{ia}\,\overline{q_{iL}}\, \eta^*\, d_{aR} + y^u_{ia}\,\overline{q_{iL}}\, \rho^*\, u_{aR} + y^D_{ij}\,\overline{q_{iL}}\, \chi^*\, D_{jR} + y^U_{i3}\,\overline{q_{iL}}\, \xi^*\, U_{3R}\\
&& +h^u_{3a}\,\overline{q_{3L}}\, \eta\, u_{aR} +
h^d_{3a}\,\overline{q_{3L}}\, \rho\, d_{aR} +
h^U_{33}\,\overline{q_{3L}}\, \chi\, U_{3R} +
h^D_{3i}\,\overline{q_{3L}}\, \xi\, D_{iR} +
h.c.~.\nn
\eea
The scalar singlet $\sigma$ and the triplet $\zeta$ do not couple to quarks at tree level. 
Moreover, even though $\xi$ couples to quarks, it does not contribute to the tree level masses as a result of matter parity conservation. 
Therefore, quark mass generation proceeds in a way similar to the minimal 331 models, where only three triplets are present {\it i.e.} $\eta, \rho$ and 
$\chi$. 

Expanding the Yukawa operators above in terms of the field components, we find that the only neutral scalars that can mediate flavour changing currents 
amongst the standard quarks are $\eta_1^0$ and $\rho_2^0$. The relevant terms are
\be 
-\mathcal{L}_q \supset  y^d_{ia}\,\overline{d_{iL}}\, (\eta_1^0)^*\, d_{aR} - y^u_{ia}\,\overline{u_{iL}}\, (\rho_2^0)^*\, u_{aR} + h^u_{3a}\,\overline{u_{3L}}\, \eta_1^0\, u_{aR} +
h^d_{3a}\,\overline{d_{3L}}\, \rho_2^0\, d_{aR} +
h.c.~.
\ee 
The CP-even and CP-odd components of $\eta_1^0$ and $\rho_2^0$ mix with other fields according to the mass matrices in Eqs. (\ref{MS2}) and (\ref{MA2}), respectively. 

Therefore, although our model introduces new scalars when compared to the minimal 331 version, the new fields ($\sigma,\zeta$ and $\xi$) 
do not imply new sources of FCNCs~\footnote{The only way some of the new scalars can contribute to tree-level FCNCs is via their mixing with $\eta_1^0$ and $\rho_2^0$.}.
This way, the results found for the 331 version in Ref. \cite{Cogollo:2012ek} can be adapted to our case. 
In that paper, the authors found that no light state at the \sm scale mediates flavour changing neutral currents. 
On the other hand, the heavy states, with masses at the 331 breaking scale $w$, do mediate FCNCs, 
hence consistency with experiment requires mediator masses of a few TeV or above.\\

We now turn to possible phenomenological implications for particle colliders.
Now that Run3 of LHC is soon starting and High-Luminosity LHC is in preparation, it is very relevant to
explore the possibility of producing the new particles in the currently planned experimental programme at CERN.


First we note that the new neutral gauge bosons present in the model yield new contributions to Drell-Yan production of di-muon events at the LHC~\cite{Aad:2019fac}.
On this basis one expects bounds at the few TeV level, as seen e.g. in Fig.~5 of Ref.~\cite{Queiroz:2016gif}. 
These are similar but complementary to the sensitivity limits obtained from meson-anti-meson mixing.

These new neutral gauge bosons also provide a portal for producing other new particles present in the model.
These include the heavy quarks with electric charge $2/3$ and $-1/3$, as well as other particles in the ``dark sector''.
From current studies on 331 models we expect that the non-observation of any signal would restrict the parameter space, giving
rise to mass limits at the few TeV level for the new particles, similar to the ones obtained above.
Hence, if the masses of the new particles are chosen to be adequately large, none of these restrictions is likely to ``kill the model''. 
Instead, a number of processes associated with well-motivated TeV-scale physics could be generated within potentially achievable experimental sensitivities.
In short, in addition to providing a comprehensive scotogenic framework for neutrino mass and dark matter, our model also provides
a rich benchmark for new physics at collider experiments.
Quantitative details require dedicated studies and simulations that lie outside the scope of this paper.

\section{Summary and conclusions}
\label{sec:Conclusions}

Here we have proposed an $\mathrm{SU(3)_c \otimes SU(3)_L \otimes U(1)_X \otimes U(1)_{N}}$ electroweak extension of the standard model where dark matter stability arises from a residual matter-parity symmetry, following naturally from the spontaneous breaking of the gauge symmetry.
The theory is scotogenic in the sense that dark matter is the mediator responsible for neutrino mass generation.
A key feature of our new scotogenic dark matter theory is the presence of a triplet scalar boson with anti-symmetric Yukawa couplings to neutrinos.
This naturally leads to a very simple characteristic prediction, i.e. one of the light neutrinos is massless, thus implying a lower bound
for the \znbb decay rate. In contrast to most other models where a massless neutrino arises from an \emph{ad hoc} incomplete multiplet choice,
here it is an unavoidable characteristic feature of the theory.
The theory also provides a comprehensive framework for scotogenic dark matter and a rich benchmark for FCNC tests and collider searches at the LHC.

\acknowledgements 
\noindent

Work supported by the Spanish grants SEV-2014-0398 and FPA2017-85216-P
(AEI/FEDER, UE), PROMETEO/2018/165 (Generalitat Valenciana) and the
Spanish Red Consolider MultiDark
FPA2017-90566-REDC. J. L. acknowledges financial support under grant
2019/04195-7, S\~ao Paulo Research Foundation (FAPESP), while OP is
supported by the National Research Foundation of Korea, under Grants
No. 2017K1A3A7A09016430 and No. 2017R1A2B4006338.
The Feynman diagram is drawn using TikZ-Feynman~\cite{Ellis:2016jkw}.


\bibliographystyle{utphys}
\bibliography{bibliography}

\providecommand{\href}[2]{#2}\begingroup\raggedright\begin{thebibliography}{10}

\bibitem{Jungman:1995df}
G.~Jungman, M.~Kamionkowski, and K.~Griest, ``{Supersymmetric dark matter},''
  \href{http://dx.doi.org/10.1016/0370-1573(95)00058-5}{{\em Phys.Rept.}
  {\bfseries 267} (1996) 195--373},
  \href{http://arxiv.org/abs/hep-ph/9506380}{{\ttfamily arXiv:hep-ph/9506380
  [hep-ph]}}.

\bibitem{deSalas:2017kay}
P.~F. de~Salas, D.~V. Forero, C.~A. Ternes, M.~Tortola, and J.~W.~F. Valle,
  ``{Status of neutrino oscillations 2018: 3$\sigma$ hint for normal mass
  ordering and improved CP sensitivity},''
  \href{http://dx.doi.org/10.1016/j.physletb.2018.06.019}{{\em Phys. Lett.}
  {\bfseries B782} (2018) 633--640},
\href{http://arxiv.org/abs/1708.01186}{{\ttfamily arXiv:1708.01186 [hep-ph]}}.

\bibitem{Boucenna:2014zba}
S.~M. Boucenna, S.~Morisi, and J.~W.~F. Valle, ``{The low-scale approach to
  neutrino masses},'' \href{http://dx.doi.org/10.1155/2014/831598}{{\em
  Adv.High Energy Phys.} {\bfseries 2014} (2014) 831598},
  \href{http://arxiv.org/abs/1404.3751}{{\ttfamily arXiv:1404.3751 [hep-ph]}}.

\bibitem{Ma:2006km}
E.~Ma, ``{Verifiable radiative seesaw mechanism of neutrino mass and dark
  matter},'' \href{http://dx.doi.org/10.1103/PhysRevD.73.077301}{{\em
  Phys.Rev.} {\bfseries D73} (2006) 077301}.

\bibitem{Hirsch:2013ola}
M.~Hirsch {\em et~al.}, ``{WIMP dark matter as radiative neutrino mass
  messenger},'' \href{http://dx.doi.org/10.1007/JHEP10(2013)149}{{\em JHEP}
  {\bfseries 1310} (2013) 149},
  \href{http://arxiv.org/abs/1307.8134}{{\ttfamily arXiv:1307.8134 [hep-ph]}}.

\bibitem{Merle:2016scw}
A.~Merle {\em et~al.}, ``{Consistency of WIMP Dark Matter as radiative neutrino
  mass messenger},'' \href{http://dx.doi.org/10.1007/JHEP07(2016)013}{{\em
  JHEP} {\bfseries 1607} (2016) 013},
  \href{http://arxiv.org/abs/1603.05685}{{\ttfamily arXiv:1603.05685
  [hep-ph]}}.

\bibitem{Alves:2016fqe}
A.~Alves {\em et~al.}, ``{Matter-parity as a residual gauge symmetry: Probing a
  theory of cosmological dark matter},''
  \href{http://dx.doi.org/10.1016/j.physletb.2017.07.056}{{\em Phys. Lett.}
  {\bfseries B772} (2017) 825--831},
\href{http://arxiv.org/abs/1612.04383}{{\ttfamily arXiv:1612.04383 [hep-ph]}}.

\bibitem{Dong:2017zxo}
P.~V. Dong {\em et~al.}, ``{The Dark Side of Flipped Trinification},''
  \href{http://dx.doi.org/10.1007/JHEP04(2018)143}{{\em JHEP} {\bfseries 04}
  (2018) 143},
\href{http://arxiv.org/abs/1710.06951}{{\ttfamily arXiv:1710.06951 [hep-ph]}}.

\bibitem{Kang:2019sab}
S.~K. Kang {\em et~al.}, ``{Scotogenic dark matter stability from gauged matter
  parity},''
\href{http://arxiv.org/abs/1902.05966}{{\ttfamily arXiv:1902.05966 [hep-ph]}}.

\bibitem{Singer:1980sw}
M.~Singer, J.~W.~F. Valle, and J.~Schechter, ``{Canonical Neutral Current
  Predictions From the Weak Electromagnetic Gauge Group SU(3) x U(1)},''
  \href{http://dx.doi.org/10.1103/PhysRevD.22.738}{{\em Phys.Rev.} {\bfseries
  D22} (1980) 738}.

\bibitem{Pisano:1991ee}
F.~Pisano and V.~Pleitez, ``{An SU(3) x U(1) model for electroweak
  interactions},'' \href{http://dx.doi.org/10.1103/PhysRevD.46.410}{{\em Phys.
  Rev.} {\bfseries D46} (1992) 410--417},
\href{http://arxiv.org/abs/hep-ph/9206242}{{\ttfamily arXiv:hep-ph/9206242
  [hep-ph]}}.

\bibitem{Frampton:1992wt}
P.~H. Frampton, ``{Chiral dilepton model and the flavor question},''
\href{http://dx.doi.org/10.1103/PhysRevLett.69.2889}{{\em Phys. Rev. Lett.}
  {\bfseries 69} (1992) 2889--2891}.

\bibitem{Hernandez:2013mcf}
A.~E. Carcamo~Hernandez, R.~Martinez, and F.~Ochoa, ``{Radiative seesaw-type
  mechanism of quark masses in $SU(3)_C \otimes SU(3)_L \otimes U(1)_X$},''
  \href{http://dx.doi.org/10.1103/PhysRevD.87.075009}{{\em Phys. Rev.}
  {\bfseries D87} no.~7, (2013) 075009},
\href{http://arxiv.org/abs/1302.1757}{{\ttfamily arXiv:1302.1757 [hep-ph]}}.

\bibitem{Boucenna:2014ela}
S.~M. Boucenna, S.~Morisi, and J.~W.~F. Valle, ``{Radiative neutrino mass in
  3-3-1 scheme},'' \href{http://dx.doi.org/10.1103/PhysRevD.90.013005}{{\em
  Phys. Rev.} {\bfseries D90} no.~1, (2014) 013005},
  \href{http://arxiv.org/abs/1405.2332}{{\ttfamily arXiv:1405.2332 [hep-ph]}}.

\bibitem{Hernandez:2014lpa}
A.~E. Carcamo~Hernandez, E.~Catano~Mur, and R.~Martinez, ``{Lepton masses and
  mixing in $SU(3)_{C}\otimes SU(3)_{L}\otimes U(1)_{X}$ models with a $S_3$
  flavor symmetry},'' \href{http://dx.doi.org/10.1103/PhysRevD.90.073001}{{\em
  Phys. Rev.} {\bfseries D90} no.~7, (2014) 073001},
\href{http://arxiv.org/abs/1407.5217}{{\ttfamily arXiv:1407.5217 [hep-ph]}}.

\bibitem{Queiroz:2016gif}
F.~S. Queiroz, C.~Siqueira, and J.~W.~F. Valle, ``{Constraining Flavor Changing
  Interactions from LHC Run-2 Dilepton Bounds with Vector Mediators},''
  \href{http://dx.doi.org/10.1016/j.physletb.2016.10.057}{{\em Phys. Lett.}
  {\bfseries B763} (2016) 269--274},
  \href{http://arxiv.org/abs/1608.07295}{{\ttfamily arXiv:1608.07295
  [hep-ph]}}.

\bibitem{Addazi:2016xuh}
A.~Addazi, J.~W.~F. Valle, and C.~A. Vaquera-Araujo, ``{String completion of an
  $\mathrm{SU(3)_c \otimes SU(3)_L \otimes U(1)_X}$ electroweak model},''
  \href{http://dx.doi.org/10.1016/j.physletb.2016.06.015}{{\em Phys. Lett.}
  {\bfseries B759} (2016) 471--478},
  \href{http://arxiv.org/abs/1604.02117}{{\ttfamily arXiv:1604.02117
  [hep-ph]}}.

\bibitem{CarcamoHernandez:2017cwi}
A.~E. Carcamo~Hernandez, S.~Kovalenko, H.~N. Long, and I.~Schmidt, ``{A variant
  of 3-3-1 model for the generation of the SM fermion mass and mixing
  pattern},'' \href{http://dx.doi.org/10.1007/JHEP07(2018)144}{{\em JHEP}
  {\bfseries 07} (2018) 144},
\href{http://arxiv.org/abs/1705.09169}{{\ttfamily arXiv:1705.09169 [hep-ph]}}.

\bibitem{Barreto:2017xix}
E.~R. Barreto, A.~G. Dias, J.~Leite, C.~C. Nishi, R.~L.~N. Oliveira, and W.~C.
  Vieira, ``{Hierarchical fermions and detectable Z' from effective
  two-Higgs-triplet 3-3-1 model},''
  \href{http://dx.doi.org/10.1103/PhysRevD.97.055047}{{\em Phys. Rev.}
  {\bfseries D97} no.~5, (2018) 055047},
\href{http://arxiv.org/abs/1709.09946}{{\ttfamily arXiv:1709.09946 [hep-ph]}}.

\bibitem{Dong:2018aak}
P.~Van~Dong {\em et~al.}, ``{Asymmetric Dark Matter, Inflation and Leptogenesis
  from $B-L$ Symmetry Breaking},''
  \href{http://dx.doi.org/10.1103/PhysRevD.99.055040}{{\em Phys. Rev.}
  {\bfseries D99} no.~5, (2019) 055040},
\href{http://arxiv.org/abs/1805.08251}{{\ttfamily arXiv:1805.08251 [hep-ph]}}.

\bibitem{Dias:2018ddy}
A.~G. Dias, J.~Leite, D.~D. Lopes, and C.~C. Nishi, ``{Fermion Mass Hierarchy
  and Double Seesaw Mechanism in a 3-3-1 Model with an Axion},''
  \href{http://dx.doi.org/10.1103/PhysRevD.98.115017}{{\em Phys. Rev.}
  {\bfseries D98} no.~11, (2018) 115017},
\href{http://arxiv.org/abs/1810.01893}{{\ttfamily arXiv:1810.01893 [hep-ph]}}.

\bibitem{Huitu:2019bvo}
K.~Huitu, N.~Koivunen, and T.~J. Karkkainen, ``{Neutrino mass via linear
  seesaw, 331-model and Froggatt-Nielsen mechanism},'' in {\em {Prospects in
  Neutrino Physics (NuPhys2018) London, United Kingdom, December 19-21, 2018}}.
\newblock 2019.
\newblock
\href{http://arxiv.org/abs/1903.12126}{{\ttfamily arXiv:1903.12126 [hep-ph]}}.
\newblock

\bibitem{Boucenna:2014dia}
S.~M. Boucenna {\em et~al.}, ``{Small neutrino masses and gauge coupling
  unification},'' \href{http://dx.doi.org/10.1103/PhysRevD.91.031702}{{\em
  Phys. Rev.} {\bfseries D91} no.~3, (2015) 031702},
  \href{http://arxiv.org/abs/1411.0566}{{\ttfamily arXiv:1411.0566 [hep-ph]}}.

\bibitem{Deppisch:2016jzl}
F.~F. Deppisch {\em et~al.}, ``{331 Models and Grand Unification: From Minimal
  SU(5) to Minimal SU(6)},''
  \href{http://dx.doi.org/10.1016/j.physletb.2016.10.002}{{\em Phys. Lett.}
  {\bfseries B762} (2016) 432--440},
  \href{http://arxiv.org/abs/1608.05334}{{\ttfamily arXiv:1608.05334
  [hep-ph]}}.

\bibitem{Reig:2016tuk}
M.~Reig, J.~W.~F. Valle, and C.~A. Vaquera-Araujo, ``{Unifying left-right
  symmetry and 331 electroweak theories},''
  \href{http://dx.doi.org/10.1016/j.physletb.2016.12.049}{{\em Phys. Lett.}
  {\bfseries B766} (2017) 35--40},
  \href{http://arxiv.org/abs/1611.02066}{{\ttfamily arXiv:1611.02066
  [hep-ph]}}.

\bibitem{Hati:2017aez}
C.~Hati {\em et~al.}, ``{Towards gauge coupling unification in left-right
  symmetric $\mathrm{SU(3)_c \times SU(3)_L \times SU(3)_R \times U(1)_{X}}$
  theories},'' \href{http://dx.doi.org/10.1103/PhysRevD.96.015004}{{\em Phys.
  Rev.} {\bfseries D96} no.~1, (2017) 015004},
\href{http://arxiv.org/abs/1703.09647}{{\ttfamily arXiv:1703.09647 [hep-ph]}}.

\bibitem{Schechter:1980gr}
J.~Schechter and J.~W.~F. Valle, ``{Neutrino Masses in SU(2) x U(1)
  Theories},'' \href{http://dx.doi.org/10.1103/PhysRevD.22.2227}{{\em
  Phys.Rev.} {\bfseries D22} (1980) 2227}.

\bibitem{Reig:2018ztc}
M.~Reig, D.~Restrepo, J.~W.~F. Valle, and O.~Zapata, ``{Bound-state dark matter
  with Majorana neutrinos},''
  \href{http://dx.doi.org/10.1016/j.physletb.2019.01.023}{{\em Phys. Lett.}
  {\bfseries B790} (2019) 303--307},
\href{http://arxiv.org/abs/1806.09977}{{\ttfamily arXiv:1806.09977 [hep-ph]}}.

\bibitem{Valle:1983dk}
J.~W.~F. Valle and M.~Singer, ``{Lepton Number Violation With Quasi Dirac
  Neutrinos},'' \href{http://dx.doi.org/10.1103/PhysRevD.28.540}{{\em
  Phys.Rev.} {\bfseries D28} (1983) 540}.

\bibitem{Ma:2014qra}
E.~Ma and R.~Srivastava, ``{Dirac or inverse seesaw neutrino masses with $B-L$
  gauge symmetry and $S_3$ flavor symmetry},''
  \href{http://dx.doi.org/10.1016/j.physletb.2014.12.049}{{\em Phys. Lett.}
  {\bfseries B741} (2015) 217--222},
\href{http://arxiv.org/abs/1411.5042}{{\ttfamily arXiv:1411.5042 [hep-ph]}}.

\bibitem{Ma:2015raa}
E.~Ma and R.~Srivastava, ``{Dirac or inverse seesaw neutrino masses from gauged
  $B–L$ symmetry},'' \href{http://dx.doi.org/10.1142/S0217732315300207}{{\em
  Mod. Phys. Lett.} {\bfseries A30} no.~26, (2015) 1530020},
\href{http://arxiv.org/abs/1504.00111}{{\ttfamily arXiv:1504.00111 [hep-ph]}}.

\bibitem{Ma:2015mjd}
E.~Ma, N.~Pollard, R.~Srivastava, and M.~Zakeri, ``{Gauge $B-L$ Model with
  Residual $Z_3$ Symmetry},''
  \href{http://dx.doi.org/10.1016/j.physletb.2015.09.010}{{\em Phys. Lett.}
  {\bfseries B750} (2015) 135--138},
\href{http://arxiv.org/abs/1507.03943}{{\ttfamily arXiv:1507.03943 [hep-ph]}}.

\bibitem{Schechter:1981bd}
J.~Schechter and J.~W.~F. Valle, ``{Neutrinoless Double beta Decay in SU(2) x
  U(1) Theories},'' \href{http://dx.doi.org/10.1103/PhysRevD.25.2951}{{\em
  Phys.Rev.} {\bfseries D25} (1982) 2951}.

\bibitem{Schechter:1980gk}
J.~Schechter and J.~W.~F. Valle, ``{Neutrino Oscillation Thought Experiment},''
\href{http://dx.doi.org/10.1103/PhysRevD.23.1666}{{\em Phys. Rev.} {\bfseries
  D23} (1981) 1666}.

\bibitem{Alduino:2017ehq}
{\bfseries CUORE} Collaboration, C.~Alduino {\em et~al.}, ``{First Results from
  CUORE: A Search for Lepton Number Violation via $0\nu\beta\beta$ Decay of
  $^{130}$Te},'' \href{http://dx.doi.org/10.1103/PhysRevLett.120.132501}{{\em
  Phys.Rev.Lett.} {\bfseries 120} (2018) 132501},
  \href{http://arxiv.org/abs/1710.07988}{{\ttfamily arXiv:1710.07988
  [nucl-ex]}}.

\bibitem{Anton:2019wmi}
{\bfseries EXO-200} Collaboration, G.~Anton {\em et~al.}, ``{Search for
  Neutrinoless Double-Beta Decay with the Complete EXO-200 Dataset},''
  \href{http://arxiv.org/abs/1906.02723}{{\ttfamily arXiv:1906.02723
  [hep-ex]}}.

\bibitem{Agostini:2018tnm}
{\bfseries GERDA} Collaboration, M.~Agostini {\em et~al.}, ``{Improved Limit on
  Neutrinoless Double-$\beta$ Decay of $^{76}$Ge from GERDA Phase II},''
  \href{http://dx.doi.org/10.1103/PhysRevLett.120.132503}{{\em Phys.Rev.Lett.}
  {\bfseries 120} (2018) 132503},
  \href{http://arxiv.org/abs/1803.11100}{{\ttfamily arXiv:1803.11100
  [nucl-ex]}}.

\bibitem{KamLAND-Zen:2016pfg}
{\bfseries KamLAND-Zen} Collaboration, A.~Gando {\em et~al.}, ``{Search for
  Majorana Neutrinos near the Inverted Mass Hierarchy Region with
  KamLAND-Zen},'' \href{http://dx.doi.org/10.1103/PhysRevLett.117.109903}{{\em
  Phys.Rev.Lett.} {\bfseries 117} (2016) 082503},
  \href{http://arxiv.org/abs/1605.02889}{{\ttfamily arXiv:1605.02889
  [hep-ex]}}.

\bibitem{Abgrall:2017syy}
{\bfseries LEGEND} Collaboration, N.~Abgrall {\em et~al.},
  \href{http://dx.doi.org/10.1063/1.5007652}{``{The Large Enriched Germanium
  Experiment for Neutrinoless Double Beta Decay (LEGEND)},''} vol.~1894,
  p.~020027.
\newblock 2017.
\newblock \href{http://arxiv.org/abs/1709.01980}{{\ttfamily arXiv:1709.01980
  [physics.ins-det]}}.

\bibitem{Andringa:2015tza}
{\bfseries SNO+} Collaboration, S.~Andringa {\em et~al.}, ``{Current Status and
  Future Prospects of the SNO+ Experiment},''
  \href{http://dx.doi.org/10.1155/2016/6194250}{{\em Adv.High Energy Phys.}
  {\bfseries 2016} (2016) 6194250},
  \href{http://arxiv.org/abs/1508.05759}{{\ttfamily arXiv:1508.05759
  [physics.ins-det]}}.

\bibitem{Albert:2017hjq}
{\bfseries nEXO} Collaboration, J.~Albert {\em et~al.}, ``{Sensitivity and
  Discovery Potential of nEXO to Neutrinoless Double Beta Decay},''
  \href{http://dx.doi.org/10.1103/PhysRevC.97.065503}{{\em Phys.Rev.}
  {\bfseries C97} (2018) 065503},
  \href{http://arxiv.org/abs/1710.05075}{{\ttfamily arXiv:1710.05075
  [nucl-ex]}}.

\bibitem{Dong:2014wsa}
P.~Dong, D.~Huong, F.~S. Queiroz, and N.~Thuy, ``{Phenomenology of the 3-3-1-1
  model},'' \href{http://dx.doi.org/10.1103/PhysRevD.90.075021}{{\em Phys.Rev.}
  {\bfseries D90} (2014) 075021},
  \href{http://arxiv.org/abs/1405.2591}{{\ttfamily arXiv:1405.2591 [hep-ph]}}.

\bibitem{Cogollo:2012ek}
D.~Cogollo, A.~de~Andrade, F.~Queiroz, and P.~Rebello~Teles, ``{Novel sources
  of Flavor Changed Neutral Currents in the $331_{RHN}$ model},''
  \href{http://dx.doi.org/10.1140/epjc/s10052-012-2029-7}{{\em Eur.Phys.J.}
  {\bfseries C72} (2012) 2029},
  \href{http://arxiv.org/abs/1201.1268}{{\ttfamily arXiv:1201.1268 [hep-ph]}}.

\bibitem{Aad:2019fac}
{\bfseries ATLAS} Collaboration, G.~Aad {\em et~al.}, ``{Search for high-mass
  dilepton resonances using 139 fb$^{-1}$ of $pp$ collision data collected at
  $\sqrt{s}=$13 TeV with the ATLAS detector},''
  \href{http://dx.doi.org/10.1016/j.physletb.2019.07.016}{{\em Phys.Lett.}
  {\bfseries B796} (2019) 68--87},
  \href{http://arxiv.org/abs/1903.06248}{{\ttfamily arXiv:1903.06248
  [hep-ex]}}.

\bibitem{Ellis:2016jkw}
J.~Ellis, ``{TikZ-Feynman: Feynman diagrams with TikZ},''
  \href{http://dx.doi.org/10.1016/j.cpc.2016.08.019}{{\em Comput. Phys.
  Commun.} {\bfseries 210} (2017) 103--123},
\href{http://arxiv.org/abs/1601.05437}{{\ttfamily arXiv:1601.05437 [hep-ph]}}.

\end{thebibliography}\endgroup
\end{document}